# Early features associated with the neurocognitive development at 36 months old: the AuBE study

Sabine Plancoulaine[a*], Camille Stagnara[b], Sophie Flori[b,c], Flora Bat-Pitault[d], Jian-Sheng Lin[e,f], Hugues Patural[b,c], Patricia Franco[e,f]

**Affiliations:**

[a] INSERM, UMR1153, Epidemiology and Statistics Sorbonne Paris Cité Research Center (CRESS), early ORigins of Child Health And Development Team (ORCHAD), Villejuif, F-94807 France; Paris-Descartes University France; [b] EA SNA-EPIS Research Laboratory 4607, Jean Monnet University, Saint-Etienne, F-42027, France; [c] Neonatal Intensive Care Unit, Department of Pediatric Medicine, CHU de Saint-Etienne, Saint-Etienne, F-42055, France; [d] Child and Adolescent Psychopathology Unit, Salvator Hospital, Public Assistance-Marseille Hospitals, Aix-Marseille II University, Marseille, F-13000, France; [e] Sleep Pediatric Unit, Woman Mother Child Hospital, Lyon1 University, F-69000, France; [f] Integrative Physiology of Brain Arousal System Research laboratory, CRNL, INSERM-U1028, CNRS UMR5292, Lyon1 University, Lyon, F-69000, France.

**Corresponding author:**

Sabine Plancoulaine, INSERM, UMR1153, Equipe 6 - ORCHAD, 16 avenue Paul Vaillant Couturier, 94800 Villejuif, France.

Phone +33145595109 - Fax +33147269454

Mail sabine.plancoulaine@inserm.fr

**Running title:** Factors associated with IQ scores at 3y

**Conflict of interest:** none

**Specific funding**: none

**Abbreviations**:
BMI – Body Mass Index;
DNR – Day/Night sleep Ratio;
FSIQ - Full Scale intelligence quotient;
HAD - Hospital Anxiety and Depression scale;
IQ - intelligence quotient;
PIQ - performance intelligence quotient;
SD – Standard Deviation;
VIQ - verbal intelligence quotient;
WPPSI-III - Weschler Preschool and Primary Scale Intelligence test

**ABSTRACT**

**Background**. Few studies on the relations between sleep quantity and/or quality and cognition were conducted among pre-schoolers from healthy general population. We aimed at identifying, among 3 years old children, early factors associated with intelligence quotient estimated through Weschler Preschool and Primary Scale Intelligence-III test and its indicators: full-scale-, verbal- and performance-intelligence quotients and their sub-scale scores.

**Methods.** We included 194 children from the French birth-cohort AuBE with both available Weschler Preschool and Primary Scale Intelligence-III scores at 3y and sleep data. Information was collected through self-questionnaires at birth, 6, 12, 18 and 24 months. A day/night sleep ratio was calculated.

**Results.** Mean scores were in normal ranges for verbal-, performance- and full-scale-intelligence quotients. In multivariate models, being a ≥3 born-child and watching television ≥1 hour/day at 24 months were negatively associated with all intelligence quotient scores while collective care arrangement was positively associated. Night waking at 6 and frequent snoring at 18 months were negatively associated with performance intelligence quotient, some subscales and full-scale-intelligence quotient contrary to day/night sleep ratio at 12 months. No association was observed between early sleep characteristics and verbal intelligence quotient.

**Conclusion.** We showed that early features including infant sleep characteristics influence intelligence quotient scores at 3 years old. Some of these may be accessible to prevention.

**Keywords:**
sleep;
intelligence quotient;
preschoolers;
cohort study;
developmental origins of health and disease (DOHaD);

**HIGHLIGHTS**

- Toddlers' early characteristics influence their cognitive scores at 3 years old
- Birth rank and TV≥1h/day at 2 decreased all scores; collective care at 2 increased them
- Night-waking at 6 and snoring at 18 months decreased performance score; day/night sleep ratio at 12 increased it
- Early sleep characteristics do not impact verbal cognitive scores
- Some early differential determinants of cognitive scores at 3 are accessible to prevention

# 1. INTRODUCTION

Lack of sleep and poor sleep quality were associated mainly in cross-sectional studies in school-aged children with poor cognitive development measured by the executive functioning, school performance, global intelligence quotient (IQ), language development, memory (working and long-term) learning, behavioural disorders (hyperactivity, aggressiveness) and attention and mood [1–3].

Very few studies were conducted among preschool children (under 6 years) from healthy general population and only eight of them were longitudinal, performed at various ages and exploring either sleep duration or sleep ratios [4–11]. Moreover, they focused mainly on language development [4,5,7], executive tasks [5,6], reasoning [6] or internalizing and externalizing problems [8]. Those studies showed that short sleep duration between 1 and 3 years old was associated with lower language [4,7] and functioning scores and with higher externalizing [8,9] and internalizing relative risks [8,10] between 2 and 6 years. Bernier et al. showed that higher night/total sleep duration ratio at 1 year old was associated with higher language scores at 2 and better reasoning at 4 years old [5], while Dionne et al showed that day/night sleep ratio at 6 and 18 months were negatively associated with language skills at 5 years old [7]. None studied the influence of fragmented sleep estimated through frequent night waking and/or frequent snoring in infancy on cognition in preschool children, and none reported the general cognitive ability measured by all Weschler Preschool and Primary Scale Intelligence (WPPSI) scales/subscales that have been shown to be highly correlated from childhood to adolescence [12]. In addition, 75% of the previously described studies were performed in North America, while sleep characteristics are different among French pre-schoolers with especially longer naps [13].

Several factors have been associated with cognitive development (e.g. preterm, maternal education, birth order)[14–16] and some are still debated, such as breastfeeding [14,17], TV viewing duration [18,19] and exposure to collective childcare [20,21].

The goal of the present study was to identify factors from early childhood influencing cognitive development in preschool children aged 36 months in a French birth-cohort with a specific emphasizing on early sleep characteristics.

# 2. METHODS
## 2.1. Study design

This secondary analysis was carried out on the French AuBE prospective mother-child cohort recruited in Saint-Etienne University Hospital between 2009 and 2011. This study first aimed at evaluating autonomic maturation profile during the first 24 months of life on psychometric development at 36 months among term and preterm new-borns. The cohort has been described elsewhere [22]. Briefly, all delivered mothers, with affiliation to a health system, were invited to participate in the first weeks post-delivery. Exclusion criteria were being less than 37 weeks corrected postnatal age and presented a documented cardiac familial history, abnormality or treatment. A total of 297 mothers and 302 children were enrolled. Follow up was 36 months. The local research ethics committee approved the study. Written informed consent was obtained at enrolment from the parents. The study has been declared in the International clinical trials registry (ClinicalTrials.gov, ID NCT00951860).

## 2.2. Data collection

### 2.2.1. Cognition at 36 months old

One psychologist administered the WPPSI third edition scale test (WPPSI-III) to all children followed up to 36 months old (N=195) avoiding test administrator variability. Children were evaluated using the five subtests as recommended for their age band (30 to 47 months): Receptive Vocabulary, Block Design, Information, Object Assembly, and Picture Naming. The whole WPPSI-III test provided subtest and composite scores that represented intellectual functioning in verbal and performance cognitive domains (further named VIQ and PIQ, respectively), and a composite score that represented the child's general intellectual ability (i.e., Full Scale IQ or FSIQ).

Other information was collected through self-questionnaires at inclusion with retrospective questions for the mother's pregnancy and prospectively at each age (6, 12, 18, 24 months) for the children.

### 2.2.2. Sleep characteristics

During the follow-up of the study, mothers completed questionnaires on their child's sleep quantity and quality at each age based on the Brief Infant Screening Questionnaire [23]. Information was also available on night- and nap- bedtimes and wake-up times from which were calculated night, day, total sleep durations per 24h and day/night sleep duration ratio (DNR) at each age. In addition, mothers were asked about sleep disturbances i.e. frequent snoring ($\geq$ 3 times a week) and night awakenings (yes/no).

### 2.2.3. Other potential influencing factors

Mothers provided at inclusion their age, the type of work that was categorized as without work, manual worker, employee, and executive worker. Mother's body mass index (BMI) before pregnancy was calculated in kg/m$^2$ based on reported weight and height and was categorized in <18.5, [18.5-25[, [25-30[, ≥30. Mothers declared parity (first-born, second-born, third and more born), term at delivery, birth weight, gender of the child. Maternal depressive symptoms were evaluated through Hospital Anxiety and Depression scale (HAD) at birth, 6, 12, 18 and 24 months. A mother was considered with depressive symptoms when HAD-depression score was ≥ 8 [24]. We collected data on breastfeeding (in weeks), type of child day care (home child minding, community child care, nursery assistant) and number of hours spent in front of screens (coded as never, <1h/day, 1 to 2 h/day, 2 and more h/day) at 24 months. To respect French ethic rules, no information was collected regarding parental origins or ethnic groups.

### 2.3. Statistical analysis

All analyses were performed using SAS software (SAS Institute 9.4, SAS Company). Student t and Chi2 tests were used to test differences between included and non-included children. Missing data for explanatory variables concerned 3.8% of the total dataset. They were imputed with median or modal value for continuous and categorical variables, respectively. Spearman's correlations were used to study correlations between continuous sleep variables. Means comparisons were performed using variance analyses. Other analyses were performed using simple and multiple linear regressions. Variables associated with a p-value <0.05 in bivariate analyses were included in the multivariate models. As we looked at early features associated with neurocognitive development, when strong correlations were observed between the same sleep variables at different ages, we kept in the model the earliest one.

## 3. RESULTS

### 3.1. Sample characteristics

One child out of the 195 children who got a WPPSI-III test at 3 years was lacking sleep information and was removed from the analyses. Thus, a total of 194 children were included. Children enrolled at birth in the AuBE cohort who were excluded of the present analysis (N=108) were more likely to have younger mother (29 versus 31 years old, p=0.001) without employment or manual workers (52% versus 23%, p<0.001). Breastfeeding duration was shorter (5 versus 9 weeks, p<0.02). No difference was identified for other variables including term at delivery, sleep characteristics, television viewing and child's care. This sample included 4 twin pairs. Their exclusion of the analysis did not change the results.

Maternal and children's general characteristics are summarized in Table 1. Mean mother's age was 31 years old (SD 4, range 19 – 42). The means scores for IQ scale at 36 months were 11 (SD 2.8, range 3 – 19) for Receptive vocabulary, 11 (SD 2.6, range 2 – 16) for Information, 11 (SD 3.0, range 5 – 19) for Picture naming, 8 (SD 2.2, range 2 – 18) for Block design, 9 (SD 2.7, range 5 – 19) for Object assembly, 106 (SD 14.4, range 62-138) for VIQ, 92 (SD 12.5, range 61-140) for PIQ and 99 (SD 13.0, range 61-138) for FSIQ. All Cronbach's coefficients alpha were >0.70. Children's sleep characteristics between 6 and 24 months are presented in Figure 1.

**Table 1.** Descriptive characteristics of the 194 children and their mother from the cohort and included in the present study.

|  | N or Means | % or SEM |
|---|---|---|
| **Sociodemographic characteristics** | | |
| **Maternal work category** | | |
| Not working | 10 | 5% |
| Manual worker | 31 | 16% |
| Employee | 135 | 70% |
| Executive worker | 18 | 9% |
| **Main child care** | | |
| Home | 39 | 20% |
| Nursery assistant | 86 | 44% |
| Community care | 69 | 36% |
| **Maternal characteristics** | | |
| **Pre-pregnancy body mass index** | | |
| <18.5 | 12 | 6% |

|  |  |  |  |
|---|---|---|---|
|  | [18.5-25[ | 136 | 70% |
|  | [25-30[ | 33 | 17% |
|  | ≥30 | 13 | 7% |
| **Smoking during pregnancy** |  |  |  |
|  | No | 158 | 81% |
|  | Yes | 36 | 19% |
| **Child characteristics** |  |  |  |
| **Term at birth** |  | 39.4 | 1.8 |
| **Breastfeeding duration (weeks)** |  | 8.7 | 11.0 |
| **Rank in the sib-ship** |  |  |  |
|  | 1st born | 79 | 41% |
|  | 2nd born | 64 | 33% |
|  | 3rd born and more | 51 | 26% |
| **TV watching at 2 years old (h/day)** |  |  |  |
|  | Never | 27 | 14% |
|  | <1 | 112 | 58% |
|  | [1-2[ | 46 | 24% |
|  | ≥2 | 9 | 5% |

**Figure 1.** Sleep characteristics among the 194 children from the cohort at 6, 12, 18 and 24 months of age. A) Mean sleep durations with total sleep duration in black, night sleep in grey and day sleep in light grey. B) Proportion of sleep troubles observed with night awakenings (yes/no) in grey, frequent snoring (≥3 times per week) in light grey. C) Day/night sleep ratio (DNR) according to age and 1 SD confidence interval.

**A.**

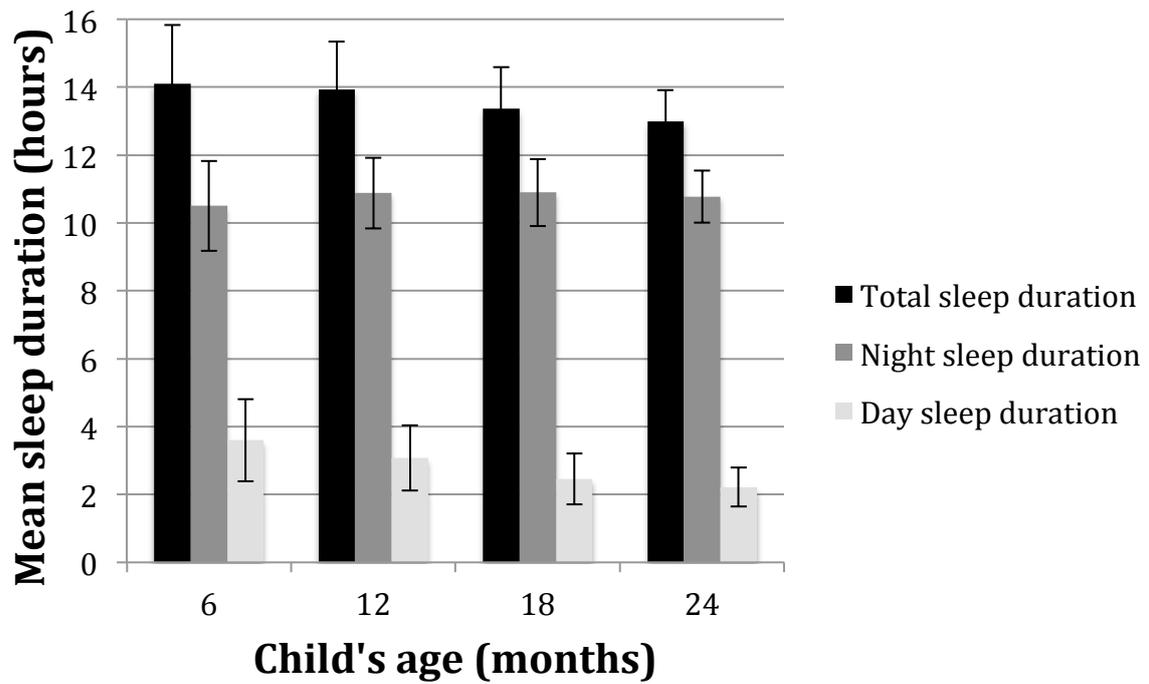

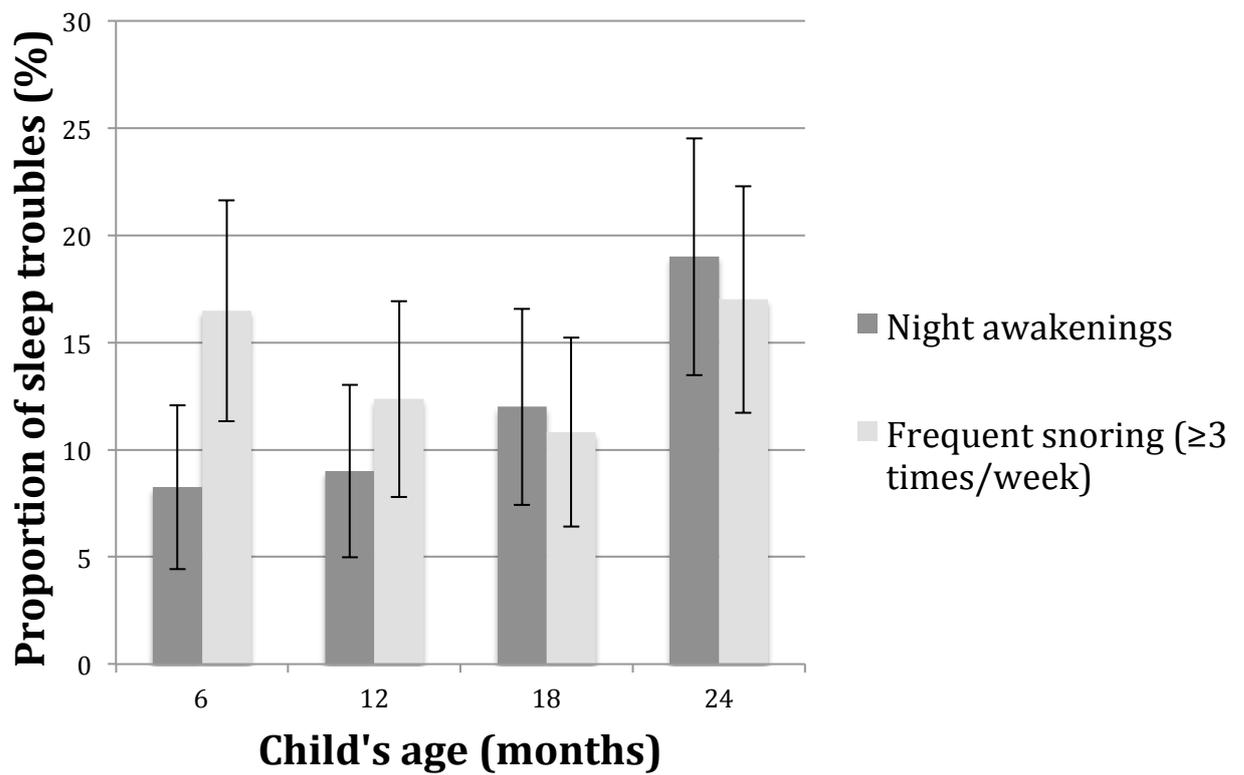

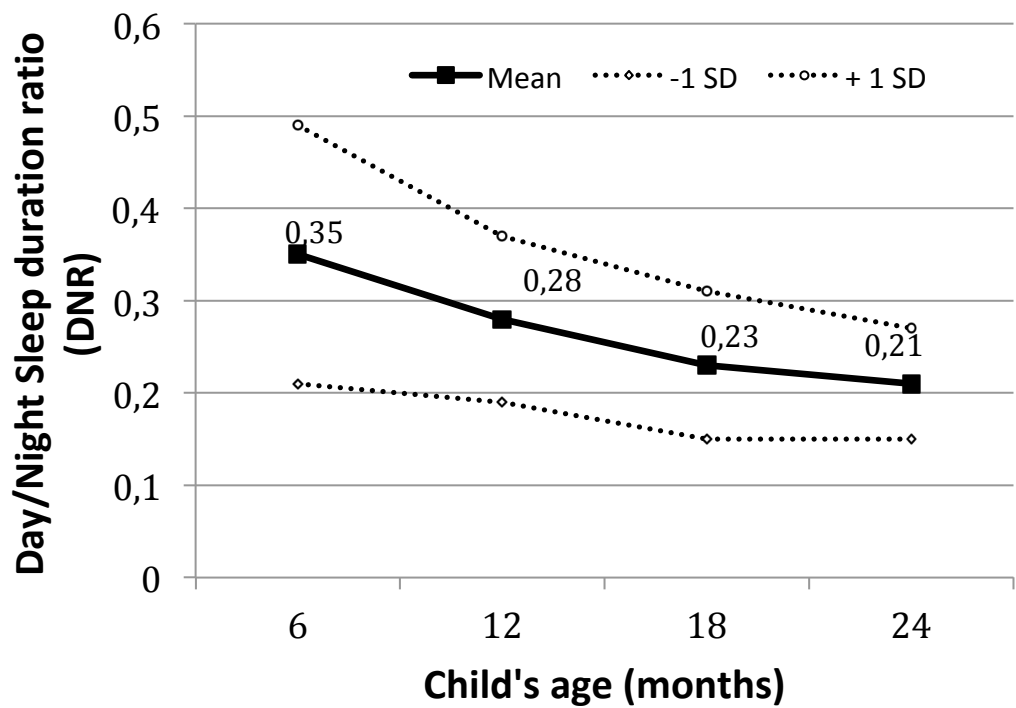

### 3.2. Early sleep factor associated with IQ scores in bivariate analyses

In bivariate analyses, sleep characteristics negatively associated with at least one IQ score were existence of night waking at 6 months (with nearly all IQ scores) and frequent snoring at

18 months (PIQ, Block Design). Sleep characteristics positively associated with at least one IQ score were DNR at 12 and 18 months (Object Assembly) and day sleep duration at 18 months (Object Assembly). There were no other associations between IQ scores and total, night or day sleep durations at any age (Table 2).

### 3.3. Correlations and associations between early sleep characteristics

No association was observed between night waking at 6 months and frequent snoring at 18 months (p=0.71). There was no correlation between night and day sleep duration at each age (p>0.80) but as expected correlations were strongly significant between day sleep duration at 12 and 18 months and between night sleep durations at 12 and 18 months ($\rho = 0.33$, $p<10^{-4}$). DNR at 12 months was strongly correlated with DNR at 18 months ($\rho = 0.29$, $p<10^{-4}$). There was no mean DNR difference for night waking at 6 months (p=0.65) or snoring at 18 months (p=0.79).

### 3.4. Multivariate model

The multivariate model showed associations between cognitive scores and socio-demographic, child general characteristics up to 24 months old (Table 3). Being a third born or more was negatively associated with cognitive scores at 36 months old, especially regarding VIQ, FSIQ, Information, Picture naming and Block design scores. However, being cared in a community centre was positively associated with all cognitive scores except receptive vocabulary and block design scores. All mean cognitive scores were negatively associated with longer time spent watching TV at 2 years old, with a dose response effect. There was no association with breastfeeding duration or maternal depressive symptoms at any age. This multivariate model also showed that early sleep characteristics were associated with cognitive scores. Night awakenings at 6 months old were associated with a decreased mean of PIQ and FSIQ scores by approximately 3.5 points and frequent snoring reported at 18 months was associated with a decreased PIQ by 7 points. DNR at 12 months was associated with an increased mean PIQ and FSIQ by 4 points.

**Table 2**. Unadjusted general and specific cognitive scores measured through Weschler Preschool and Primary Scale Intelligence (WPPSI)-III at 36 months old and child sleep characteristics between 6 months and 24 months old among the 194 children from the cohort. General scores measured are Verbal, Performance and Full Scale IQs; Specific scores measured are Receptive Vocabulary, Information, Picture naming, Block design and Object Assembly scores. Day/Night sleep duration Ratio (DNR) is per 0.1 unit.

| | Verbal IQ β (SD) | Performance IQ β (SD) | Full Scale IQ β (SD) | Receptive vocabulary β (SD) | Information β (SD) | Picture Naming β (SD) | Block Design β (SD) | Object Assembly β (SD) |
|---|---|---|---|---|---|---|---|---|
| **Sleep characteristics at 6 months** | | | | | | | | |
| **Night awakenings** | | | | | | | | |
| No | - | - | - | - | - | - | - | - |
| Yes | -2.44 (2.08) | -4.34 (1.79)* | -3.81 (1.87)* | -0.37 (0.40) | -0.43 (0.38) | -0.92 (0.42)* | -0.75 (0.32)* | -0.67 (0.39) |
| **Frequent snoring** | | | | | | | | |
| No | - | - | - | - | - | - | - | - |
| Yes | 1.23 (2.80) | 1.27 (2.43) | 1.24 (2.53) | 0.05 (0.54) | 0.30 (0.51) | 0.13 (0.57) | 0.35 (0.43) | 0.05 (0.53) |
| **Day sleep duration** | -0.69 (0.86) | -1.02 (0.74) | -1.04 (0.77) | -0.03 (0.17) | -0.23 (0.16) | -0.29 (0.18) | -0.19 (0.13) | -0.14 (0.16) |
| **Night sleep duration** | 1.00 (0.78) | 0.73 (0.68) | 1.05 (0.71) | 0.20 (0.15) | 0.17 (0.14) | 0.26 (0.16) | 0.11 (0.12) | 0.14 (0.15) |
| **Total sleep duration** | 0.24 (0.60) | -0.07 (0.52) | 0.10 (0.54) | 0.10 (0.12) | -0.01 (0.11) | 0.01 (0.12) | -0.03 (0.09) | 0.01 (0.11) |
| **DNR** | -1.03 (0.80) | -1.18 (0,65) | -1.33 (0.68) | -0.12 (0.15) | -0.26 (0.14) | -0.33 (0.15) | -0.22 (0.16) | -0.16 (0.14) |
| **Sleep characteristics at 12 months** | | | | | | | | |
| **Night awakenings** | | | | | | | | |
| No | - | - | - | - | - | - | - | - |
| Yes | -1.57 (2.07) | 0.89 (1.80) | -0.46 (1.88) | -0.37 (0.40) | -0.22 (0.38) | -0.71 (0.42) | -0.37 (0.32) | 0.7 (0.39) |
| **Frequent snoring** | | | | | | | | |
| No | - | - | - | - | - | - | - | - |
| Yes | 1.46 (3.15) | 2.52 (2.73) | 2.43 (2.85) | -0.14 (0.61) | 0.64 (0.57) | 0.44 (0.65) | 0.01 (0.49) | 0.81 (0.59) |
| **Day sleep duration** | 0.72 (1.08) | 0.85 (0.94) | 0.92 (0.98) | 0.17 (0.21) | 0.06 (0.20) | 0.07 (0.22) | 0.00 (0.17) | 0.29 (0.20) |
| **night sleep duration** | 0.02 (1.00) | -1.05 (0.87) | -0.62 (0.90) | -0.08 (0.19) | 0.08 (0.18) | 0.04 (0.20) | -0.04 (0.15) | -0.32 (0.19) |
| total sleep duration | 0.34 (0.74) | -0.18 (0.64) | 0.09 (0.67) | 0.04 (0.14) | 0.07 (0.13) | 0.05 (0.15) | -0.02 (0.11) | -0.04 (0.14) |
| DNR | 0.66 (1.11) | 1.21 (0.96) | 1.1 (1.00) | 0.19 (0.21) | 0.02 (0.20) | 0.04 (0.23) | 0.02 (0.17) | 0.40 (0.21)* |
| **Sleep characteristics at 18 months** | | | | | | | | |
| **Night awakenings** | | | | | | | | |
| No | - | - | - | - | - | - | - | - |
| Yes | 0.32 (2.10) | -0.75 (1.83) | -0.2 (1.90) | 0.09 (0.41) | -0.06 (0.38) | 0.04 (0.43) | -0.07 (0.33) | -0.12 (0.40) |
| **Frequent snoring** | | | | | | | | |
| No | - | - | - | - | - | - | - | - |
| Yes | 0.42 (3.34) | -5.32 (2.88)* | -2.70 (3.01) | -0.43 (0.64) | 0.62 (0.61) | 0.75 (0.68) | -1.00 (0.51)* | -0.69 (0.63) |
| Day sleep duration | 1.11 (1.38) | 2.04 (1.19) | 1.87 (1.24) | 0.26 (0.27) | 0.14 (0.25) | 0.47 (0.28) | 0.02 (0.21) | 0.67 (0.26)* |
| night sleep duration | 0.78 (1.05) | -0.24 (0.92) | 0.36 (0.95) | 0.11 (0.20) | 0.18 (0.19) | 0.12 (0.22) | 0.06 (0.16) | -0.14 (0.20) |
| total sleep duration | 0.91 (0.84) | 0.61 (0.73) | 0.93 (0.76) | 0.17 (0.16) | 0.17 (0.15) | 0.25 (0.17) | 0.05 (0.13) | 0.16 (0.16) |
| DNR | -0.01 (1.35) | 1.46 (1.17) | 0.82 (1.22) | 0.11 (0.26) | -0.13 (0.25) | 0.25 (0.28) | -0.10 (0.21) | 0.58 (0.25)* |
| **Sleep characteristics at 24 months** | | | | | | | | |
| **Night awakenings** | | | | | | | | |
| No | - | - | - | - | - | - | - | - |
| Yes | -1.36 (2.08) | -3.55 (1.79) | -2.87 (1.87) | -0.35 (0.40) | -0.15 (0.38) | -0.41 (0.42) | -0.72 (0.32)* | -0.42 (0.39) |
| **Frequent snoring** | | | | | | | | |
| No | - | - | - | - | - | - | - | - |
| Yes | -2.35 (2.76) | -2.42 (2.40) | -2.74 (2.49) | -0.61 (0.53) | -0.15 (0.50) | -0.10 (0.57) | -0.32 (0.43) | -0.47 (0.52) |
| Day sleep duration | 0.45 (1.82) | 1.32 (1.58) | 1.02 (1.64) | 0.33 (0.35) | -0.15 (0.33) | -0.11 (0.37) | 0.38 (0.28) | 0.05 (0.35) |
| night sleep duration | -1.60 (1.35) | -0.97 (1.18) | -1.50 (1.22) | -0.22 (0.26) | -0.32 (0.25) | -0.55 (0.28) | -0.21 (0.21) | -0.13 (0.26) |
| total sleep duration | -0.94 (1.13) | -0.16 (0.98) | -0.64 (1.02) | -0.03 (0.22) | -0.28 (0.20) | -0.42 (0.23) | 0.00 (0.18) | -0.07 (0.21) |
| DNR | 0.48 (1.83) | 1.45 (1.59) | 1.13 (1.65) | 0.26 (0.35) | -0.07 (0.33) | 0.03 (0.38) | 0.41 (0.28) | 0.07 (0.35) |

* p-value <0.05

**Table 3.** Adjusted general and specific cognitive scores measured through Weschler Preschool and Primary Scale Intelligence (WPPSI)-III at 36 months old and child sleep characteristics between 6 months and 18 months old adjusted on socio-demographic, maternal and child's characteristics among the 194 children from the cohort. General scores measured are Verbal, Performance and Full Scale IQs; Specific scores measured are Receptive Vocabulary, Information, Picture naming, Block design and Object Assembly scores. Day/Night sleep duration Ratio (DNR) is per 0.1 unit.

|  | Verbal IQ β (SD) | Performance IQ β (SD) | Full Scale IQ β (SD) | Receptive vocabulary β (SD) | Information β (SD) | Picture Naming β (SD) | Block Design β (SD) | Object Assembly β (SD) |
|---|---|---|---|---|---|---|---|---|
| **Sociodemographic characteristics** | | | | | | | | |
| **Maternal work category** | | | | | | | | |
| Without work | -6.39 (4.85) | 7.16 (4.06) | 0.48 (4.14) | -1.07 (0.98) | -1.26 (0.86) | -0.52 (1.00) | 0.98 (0.75) | 0.74 (0.96) |
| Manual worker | 0.60 (2.89) | -3.99 (2.42) | -1.65 (2.47) | 0.52 (0.59) | -0.23 (0.52) | 0.25 (0.59) | -0.60 (0.45) | -0.47 (0.57) |
| Employee | - | - | - | - | - | - | | |
| Executive worker | 0.07 (3.52) | 3.43 (2.94) | 2.44 (3.00) | -0.58 (0.71) | 0.86 (0.63) | 0.46 (0.72) | 0.23 (0.53) | 0.80 (0.68) |
| **Main child care** | | | | | | | | |
| Home | - | - | - | - | - | - | | |
| Nursery assistant | 1.95 (3.05) | 3.06 (2.55) | 3.11 (2.60) | 0.73 (0.62) | 0.06 (0.54) | 0.36 (0.63) | 0.31 (0.47) | 0.69 (0.60) |
| Community care | 6.72 (3.02)* | 7.27 (2.53)** | 8.34 (2.58)** | 1.17 (0.61) | 1.15 (0.54)* | 1.46 (0.62)* | 0.82 (0.47) | 1.26 (0.60)* |
| **Maternal characteristics** | | | | | | | | |
| **Pre-pregnancy BMI** | | | | | | | | |
| <18.5 | 5.67 (4.17) | -0.18 (3.49) | 2.82 (3.56) | 0.95 (0.85) | 0.86 (0.74) | 1.36 (0.86) | 0.83 (0.65) | -0.99 (0.83) |
| [18.5-25[ | - | - | - | - | - | - | | |
| [25-30[ | -3.97 (2.78) | -3.54 (2.32) | -4.05 (2.37) | -0.37 (0.56) | -0.87 (0.49) | -1.44 (0.57)* | -0.40 (0.43) | -0.34 (0.55) |
| ≥30 | -2.62 (3.95) | -1.12 (3.31) | -1.98 (3.37) | -0.05 (0.80) | -0.78 (0.70) | -0.53 (0.81) | 0.68 (0.61) | -1.13 (0.79) |
| **Smoking during pregnancy** | | | | | | | | |
| No | - | - | - | - | - | - | - | - |
| Yes | 0.43 (2.54) | -2.01 (2.12) | -0.79 (2.16) | 0.13 (0.51) | 0.00 (0.45) | 0.37 (0.52) | -0.95 (0.39) | 0.14 (0.50) |
| **Child characteristics** | | | | | | | | |
| **Term at birth** | 0.69 (0.60) | 0.88 (0.50) | 0.87 (0.51) | 0.15 (0.12) | 0.07 (0.11) | 0.15 (0.12) | 0.13 (0.09) | 0.20 (0.12) |
| **Birth rank** | | | | | | | | |
| 1st born | - | - | - | - | - | - | | |
| 2nd born | -0.06 (2.43) | -0.91 (2.03) | -0.38 (2.07) | -0.1 (0.49) | 0.15 (0.43) | -0.17 (0.50) | -0.05 (0.37) | 0.01 (0.47) |
| 3rd born and more | -6.54 (2.55) | -4.32 (2.13) | -6.12 (2.18)** | -1. (0.52) | -1.07 (0.45)* | -1.40 (0.52)** | -0.88 (0.39)* | -0.34 (0.50) |
| **TV watching at 2 years old (h/day)** | | | | | | | | |
| Never | -0.69 (2.91) | -2.86 (2.43) | -1.91 (2.48) | -0.37 (0.59) | 0.16 (0.52) | -0.25 (0.60) | -0.65 (0.45) | -0.61 (0.57) |
| <1 | - | - | - | - | - | | | |
| [1-2[ | -6.00 (2.55)* | -5.83 (2.14)** | -6.72 (2.18)** | -0.84 (0.52) | -1.17 (0.45)* | -0.71 (0.52) | -1.18 (0.39)** | -0.70 (0.50) |
| ≥2 | -14.41 (4.68)** | -10.29 (3.91)** | -13.86 (3.99)*** | -2.34 (0.95)* | -2.39 (0.83)** | -1.66 (0.96) | -1.52 (0.73)* | -1.92 (0.93)* |
| **Sleep characteristics** | | | | | | | | |
| **Night awakenings at 6 months** | | | | | | | | |
| No | - | - | - | - | - | - | - | - |
| Yes | -2.06 (1.98) | -3.89 (1.65)* | -3.34 (1.69)* | -0.38 (0.40) | -0.29 (0.35) | -0.85 (0.41)* | -0.71 (0.30) | -0.56 (0.38) |
| **DNR at 12 months** | 1.55 (1.10) | 2.45 (0.92)** | 2.30 (0.93)* | 0.27 (0.22) | 0.21 (0.20) | 0.27 (0.22) | 0.24 (0.17) | 0.59 (0.21)** |
| **Frequent snoring at 18 months** | | | | | | | | |
| No | - | - | - | - | - | - | - | - |
| Yes | -0.78 (3.21) | -7.03 (2.68)** | -4.39 (2.73) | -0.43 (0.65) | 0.23 (0.57) | 0.39 (0.66) | -1.24 (0.48)** | -1.04 (0.62) |

\* p-value <0.05  
\*\* p-value <0.01

## 4. DISCUSSION

### 4.1. Early sleep characteristics associated with neurocognitive development at 36 months

We showed for the first time an association between night awakenings reported by parents at 6 months and lower performance and general cognitive scores at 36 months old, independently of birth rank. One of the reasons for night awakenings at 6 months may be breastfeeding that was recently associated among French pre-schoolers with both higher scores for language skills at 2 years old [17] and with favourable changes in language skills between 2 and 3 years old [25]. The present study showed no association between breastfeeding duration or prevalence and IQ scores. Moreover, there was no difference of breastfeeding duration or prevalence between children with or without night awakenings at 6 months, suggesting that parents may only have reported "problematic" night awakenings and not awakenings for breastfeeding if any during the night. Fragmented sleep has been associated with memory difficulties and poorer academic achievement in older children [2] probably due to lower sleep efficiency tracking throughout childhood [26]. Thus, sleep was described as powerful aid in memory consolidation [27] while infants were shown to retain information for shorter periods of time than older children [28] suggesting an important role for naps.

Sleep and nap play indeed an essential role, as shown by experimental studies, in establishing short and long-term memory among toddlers and preschool children. Gómez et al showed that 15 months old infants who napped were able to abstract the general grammatical pattern of a briefly presented artificial language [29]. More particularly, those who napped within 4 hours after artificial language learning remembered the general grammatical pattern 24h later while those who did not nap showed no evidence of remembering anything [30]. More recently Giganti et al reported that nap benefits in explicit memory consolidation among children aged 3 to 6 years old but not in implicit perceptual learning (naming pictures) [31]. Kurdziel et al reported that naps help in memorization and learning especially regarding visuospatial task (Memory game) among pre-schoolers aged between 36 and 67 months, with long term benefits on memory consolidation and benefit of the nap relative to wake greatest for children who napped habitually [32]. When considering naturally occurring sleep, Lukowski et al showed that nap duration was positively associated with both immediate imitation encoding, particularly encoding in a correct temporal order and with delayed recall in generalization of temporal order information, in 10 months infants, while night-time sleep duration was not associated with immediate or delayed recalls [33]. These latter results are consistent with ours regarding DNR and performance, object assembly and full scale IQ scores. They indeed suggest an important role for sleep repartition on 24h and specifically for daytime sleep to consolidate performance memory at 12 months old.

Our results differed from those reported by the only two other publications on sleep ratios among pre-schoolers [6,7]. The first one showed, among 65 children, a positive association between night/total sleep ratio at 12 months old and executive functioning (Matrix Reasoning subscale from WPSSI III) at 4 years, taking into account previous socioeconomic status and prior cognitive functioning measured at 1 and 2 years old. The second one showed, among 1029 twins, negative associations between DNR at 6 and 18 months and language skills at 30 and 60 months with a highly heritable DNR at 6 months and more shared environmental influences on DNR at 18 months. Evaluations were performed through CDI and PPVT. They used structural equations and included child (gender, term and weight at birth, Apgar score, hospital stay duration, difficult temperament) and mother's (education, family income, depressive symptoms, overprotectiveness and impact perception) variables. DNR could be considered as a good measure of sleep consolidation during early childhood [7]. A recent systematic review of the effects of napping on measures of child development and health [34] showed that increased nap duration was associated with decreased night sleep duration and quality. This negative correlation could be implicated in the negative effect found between high day/night ratio at 6- and 18-month and language skills at 60 months [7] or higher night/total sleep ratio at 1 year and better executive performances at 4 years [6]. In our studied population, even if we found a tendency for a decrease in daytime sleep between 6 and 24 months, there was no change in the night sleep durations during this period (Figure 1 A). We could not explain why these children had longer daytime sleep. However, the positive effect of daytime sleep on learning [29–33], emotional regulation [35] and behavioral development in childhood [36] have already been reported by numerous authors.

Independently of night awakenings signalled to parents, frequent snoring at 18 months was associated with lower PIQ and Block design scores. Snoring is a commonly reported symptom of obstructive sleep apnoea, responsible for fragmented sleep, that was associated in school-aged children with greater attention deficit, reduced IQ scores, low school performance and memory deficit [37]. Limited studies have investigated the effect of habitual snoring on development of preschool children [38,39] and fewer considered toddlers [40,41]. However, they all showed lower cognitive development in children who snored even if they did not present episode of obstructive hypopnea or apnoea.

**4.2. Other factors associated with neurocognitive development at 36 months**

This study confirms among pre-schoolers the association of factors already shown associated in neurocognitive development among school-aged children as sib-ship rank and some modifiable ones as childcare arrangement and TV viewing duration.

We found a negative association between birth rank and VIQ, FSIQ, Information, Picture naming, and Block design scores. These results are consistent with previous publications suggesting that the proportion of parent-child interaction decreased with the increased of sib-ship size [42], especially regarding language development [16].

Formal structured childcare was associated with higher cognitive scores especially for language among pre-schoolers and mathematical achievement among school-aged children [20,43]. We here showed strong associations between community childcare at 24 months and all composite IQ scores (VIP, PIQ and FSIQ) at 36 months but more specifically with the Information, Picture Naming and Object Assembly scores, independently of socioeconomic status approximated through maternal work category. Formal day-care has been shown to particularly improve cognitive scores of children from families with low incomes and low parental education [43]. The children followed up to 3 years old in the present study were from medium-high socio economic level (79% ≥ employees) suggesting a positive effect even in this advantaged population.

Television viewing has been suggested to be particularly detrimental to cognitive performances and language development among school-aged children [18]. Few studies were performed among preschool children, all focusing on language development and showing controversial results [18,19,44]. In particular, Schmidt et al reported the lack of association between mean television viewing duration between birth and 2 years old and PPVT scores at 3 years old, this latter test being highly correlated with receptive vocabulary score of the WPPSI-III test. In the present study, while TV viewing duration negatively impacted all scores, the receptive vocabulary score was one of the verbal components less impacted by [19].

### 4.3. Limitations

Our study included a small sample size. This should lead to lack of power to detect an effect. However, we identified strong effects, stable across analyses and models. Due to the recruitment method, based on voluntary participation in the study, and to the attrition during follow-up, there is over-representation of middle-high working class mothers, compared to the national population, limiting the generalization of our results. Most of the data were collected through auto-questionnaires that may lead for example to either under or over estimation of sleep characteristics. The main aim of the cohort was not to study sleep and systematic bias for these specific variables seems weak. Objective assessment of sleep through recurrent sleep recording would have reduced this possibility. These were secondary analyses and only partial information was available regarding some potential risk factors. No information on parental origins or ethnic group was collected while several factors may be influenced by ethnic differences as socioeconomic and educational levels or television viewing duration [45]. No formal information on maternal IQ or educational level was available and we approximated them with work category. We did not measure the content of television programs viewed by the children (e.g. educational, recreational, violent) neither the timing of television viewing (before nap- or night-sleep) that have been shown to influence cognition [46,47].

### 4.4. Conclusion

This study conducted in French preschool children showed that in addition to modifiable habits such as television viewing duration and childcare arrangement early sleep characteristics specifically influenced IQs and subscales scores at 36 months old. Further studies are needed to confirm these results in larger samples of children from general healthy population and to examine whether these modifications track throughout childhood.

## 5. ACKNOWLEDGMENTS

The AuBE study is allowed through consecutive grants from the French Ministry of Health: *Programmes Hospitaliers de Recherche Clinique* – PHRC interregional, 2009 and AOL 2010. The neuropsychological evaluation and the study were made possible through grants from laboratories Abbott-France, Nestlé-France and the association ADERPS (Saint-Etienne non-profit organization for the pediatric research development – University Hospital of Saint-Etienne – France).

We especially acknowledge Mrs Sophie Foucat and all the parents of the association SA VIE (www.sa-vie.fr; Nantes-France), Dr Elisabeth Briand-Huchet and Mrs Myriam Morinay, president of the French national non-profit organization NAITRE et VIVRE (naitre-et-vivre.org) for supporting the development and funding of research about the Sudden Infant Death Syndrome.

We finally thank the parents and children of the AuBE cohort without whom this great adventure could not go on.